\documentclass[aps,twocolumn,bibnotes]{revtex4}
\usepackage{amsmath}

\begin{document}

\title{Comments on `Stable Quantum Computation of Unstable Classical Chaos,'\\ `Efficient Quantum Computing Insensitive to Phase Errors,'\\ and `Quantum Computer Inverting Time Arrow for Macroscopic Systems'}

\author{Alec \surname{Maassen van den Brink}}
\email{alec@dwavesys.com}
\affiliation{D-Wave Systems Inc., 320-1985 West Broadway, Vancouver, BC, V6J 4Y3, Canada}

\date{\today}

\maketitle

Recently, a Letter~\cite{GS1} published by Georgeot and Shepelyansky has been criticized~\cite{zalka,diosi}. The Letter claims an exponential speedup and reduction in error sensitivity, when phase-space density evolution under the Arnold cat map is performed on a quantum computer (QC). On the one hand, some points have not yet been made in~\cite{zalka,diosi}. On the other, the authors' reaction~\cite{GS3} raises new issues. The present note addresses both.

The first point concerns stability of the classical and quantum algorithms under their respective ``natural errors.'' By now, a consensus seems established that such comparison is fair only if both contestants operate on a discretized $N\times N$ map, say $N=2^n$. Then, the quantum errors considered are unitary gate noise (amplitude errors), plus phase errors either in those gates or during storage. Classical errors are taken to be one-cell shifts. The problem here is not a fine point of QC theory, but a misconstruction of \emph{classical} binary numbers. In fact, errors in all bits of a register are equally likely, and digital technology is based on getting all of them right most of the time, so that integers are processed error-free. This is what enabled the simulations in~\cite{GS1}. Thus, the classical algorithm has been deliberately handicapped.

Further, consider the evaluation of high harmonics of the evolved density, Fig.~1 in~\cite{GS3}. Like Fig.~1 in~\cite{GS1}, obtained on a \emph{simulated} QC, it offers a glimpse of a real one's wave function which nature does not allow one to see. Physically obtainable is a quantum measurement of the Fourier vector~$\vec{k}$, yielding nontrivial information if the $\vec{k}$-density has strong peaks at high~$k$. This is indeed the case for the cat map (\cite{GS3}, Fig.~1, bottom left and top right). Unfortunately, the latter's evolution closes in $\vec{k}$\nobreakdash-space~\cite{GS3}, and can be handled classically without effort. In contrast, the data shown (middle/bottom right) for a nontrivial ``perturbed cat map'' only have a slanted band of density---in all likelihood a mere transient due to the low number of iterations $t=5$. The quantum approach now yields the gross features of this band. Fine structure, e.g.\ the small-amplitude low-$k$ peaks for the original map, is much easier obtained classically. Thus, the implicit claim (\cite{GS3}, bottom of col.~1) of being in the same situation as Shor's algorithm (searching \emph{one} period, verifiable a posteriori, of a spiked 1D $k$-space distribution) is valid only for a narrow set of questions, not substantiating the authors' original scope.

Given the chronology, it is remarkable that Refs.~\onlinecite{zalka,diosi} do not consider the follow-up~\cite{GS2}. The latter, cited in the authors' defence~\cite{GS3}, boldly concludes that ``the massive parallelism of QC is not necessarily related to quantum interference.'' Again, one is shown figures of inaccessible wave functions, involving a phase-space coarse-graining of which no benefit or application is given except that it can apparently be performed quickly on a QC. It is worthwhile to read the ``more detailed discussion of measurements'' in version~2 of~\cite{GS2}. The added penultimate paragraph, especially a casual remark in its last sentence, in fact refutes the whole paper and indeed its very title. An \emph{in}coherent QC turns out to be an---expensive---classical one with random initial conditions, a basic point much in line with the remarks in~\cite{diosi}.

The discussion of~\cite{GS4}, also cited in~\cite{GS3}, can be brief: the only research added is an extension of the map's $y$\nobreakdash-axis, i.e., all the critiques~\cite{zalka,diosi} apply. Note that the QC derives its memory capacity from superposition. Thus, the evolution of an ensemble of orbits is necessarily \emph{linear}, corresponding to chaotic one-particle dynamics rather than to a classical gas with collisions. Hence, calling the problem ``macroscopic'' is an exaggeration, and Boltzmann is better left out of it.

In conclusion, the authors continue an unreasonably positive appraisal of their QC algorithm vis-a-vis the classical one. In particular, Ref.~\onlinecite{GS2}, instead of supporting the authors' case, is at fault in its basic premise and should be withdrawn.

I thank C.W.J. Beenakker, Yu.V. Gusev, P. Silvestrov, and  A.M. Zagoskin for discussions.

\end{document}